\documentclass[11pt,a4paper]{article}
\usepackage{eurosym}
\usepackage{amssymb}
\usepackage{amsfonts}
\usepackage{jcappub,natbib}

\pdfoutput=1
\affiliation{Pushkov Institute of Terrestrial Magnetism, Ionosphere and Radiowave
Propagation of the Russian Academy of Science (IZMIRAN), \\
108840 Troitsk, Moscow, Russia}
\emailAdd{leinson@yandex.ru}
\abstract{The observed anomalous steady decrease in surface temperature of the supernova remnant Cassiopeia A (Cas A), which was reported about ten years ago, has generated much debate. Several exotic cooling scenarios have been proposed using non-standard assumptions about the physics and evolution of this neutron star (NS). At present, significant corrections have been made to the observational data, which make it possible to numerically simulate the Cas A NS cooling process in the framework of the scenario of minimal neutrino cooling. If there is an additional source of cooling, such as axion emission, the steepness of the Cas A NS surface temperature drop will increase with the growth of the axion-nucleon interaction strength. This makes it possible to limit the minimum value of the axion decay constant $f_a$ using the condition that the NS surface temperature should be within  the 99\% confidence interval obtained from the observational data. Two types of axion models are considered: the Kim-Shifman-Weinstein-Zakharov -- KSVZ model and the Dean-Fischler-Srednitsky-Zhitnitsky --DFSZ model.
 The above criterion gives a lower limit on the axion decay constant, $f_a>3\times 10^7$ GeV and $f_a>4.5\times 10^8$ GeV  for KSVZ and DFSZ axions, respectively.}
\keywords{neutron stars, axions, supernova neutrinos}

\begin{document}

\title{Impact of axions on the Cassiopea A neutron star cooling.}
\author{Lev B. Leinson}
\maketitle

\flushbottom

\section{Introduction}

\label{sec:intro}

Cooling isolated neutron stars are of exceptional interest to
astrophysicists, because these extremely dense compact objects in the
Universe serve as a kind of laboratory for the study of matter, which cannot
be reproduced in laboratory conditions. Studying thermal evolution of
isolated neutron stars in X-rays is of a great importance for better
understanding the evolution of such objects and provides a possibility to
investigate their composition and structure (see e.g., \cite{P04,P09,YP04}).
However, it is difficult to determine the actual thermal radiation from the
surface, since the magnetosphere can emit non-thermal X-rays. Therefore, it
seems to be a great success if, among the many observed neutron stars, an
object is found whose X-ray radiation can be unambiguously associated with
the temperature of its surface.

In this regard, over the past two decades, much attention has been paid to
the thermal X-ray emission of a neutron star (NS) at the center of the
Cassiopeia A (Cas A) supernova remnant\footnote{%
The supernova remnant in Cassiopeia A contains a young ($\approx 340$ yr old 
\cite{F06}) neutron star which was discovered by Chandra satellite \cite%
{Hu00} in 1999.}. About ten years ago, \cite{H09,H10} analyzed data from
Chandra's observations over a decade and reported an anomalous steady
decrease in surface temperature $T_{s}$ by about 4\%, which they interpreted
as a direct observation of the cooling of NS Cas A, a phenomenon that had
never been observed for any isolated NS.

We shall discuss later the current state of these observations, at the
moment we note that although the real cooling rate is under debate one can
not exclude that the Cas A NS cooling is extraordinarily fast. Such a rapid
drop in surface temperature (if it occurs) is in conflict with standard
cooling scenarios based on the efficient modified Urca process. If the NS in
Cas A underwent standard cooling (through neutrino emission from the core
due to the modified Urca process) its surface temperature decline in 10
years would be $0.2\%-0.3\%$ \cite{Y01,p06}.

A rapid decrease, but a relatively high surface temperature (about $2\times
10^{6}$ K) requires a sharp change in the properties of neutrino emission
from NS. Some exotic cooling scenarios have been proposed using non-standard
assumptions about the physics and evolution of NS, including softened pion
modes \cite{B12}, quarks \cite{s13,n13} or cooling after heating process in
r-mode \cite{y11}. The existence of softened pions or quarks in the NS core
depends mainly on the density of the substance, but not on the temperature.
If this rapid cooling were constant since the birth of the NS, the current
temperature would have to be much lower than currently measured.

It is reasonable to assume \cite{P,St} that the cooling was initially slow,
but that it later accelerated significantly. In this case, the rapid
decrease in temperature can naturally be explained in terms of the minimal
cooling paradigm \cite{P04,P09}, which assumes that the rapid cooling of a
neutron star is caused by neutron superfluidity in the core. This scenario
assumes that neutrons have recently become superfluid (in the $^{3}$P$_{2}$
triplet state) in the NS core, causing a huge neutrino flux as a result of
thermal Cooper pair breaking and formation (PBF) processes that accelerates
the cooling of \cite{P,St}, while the protons were already in the
superconducting singlet state $^{1}$S$_{0}$ with a higher critical
temperature. Although this mechanism is consistent with the generally
accepted cooling paradigm, theoretical simulations have shown \cite{St,E13}
that PBF processes in neutron triplet condensate are not efficient enough to
explain the rapid temperature drop. This stimulated the work of \cite{L14},
where axion emission was added to compensate for the deficit of neutrino
energy losses from the Cas A NS and reproduce the seeming rapid cooling of
this object reported in \cite{H09,H10}.%%%%%%%%%%%%%%%%%%
\begin{figure}[tbp]
\begin{center}
\includegraphics[width =1\textwidth]{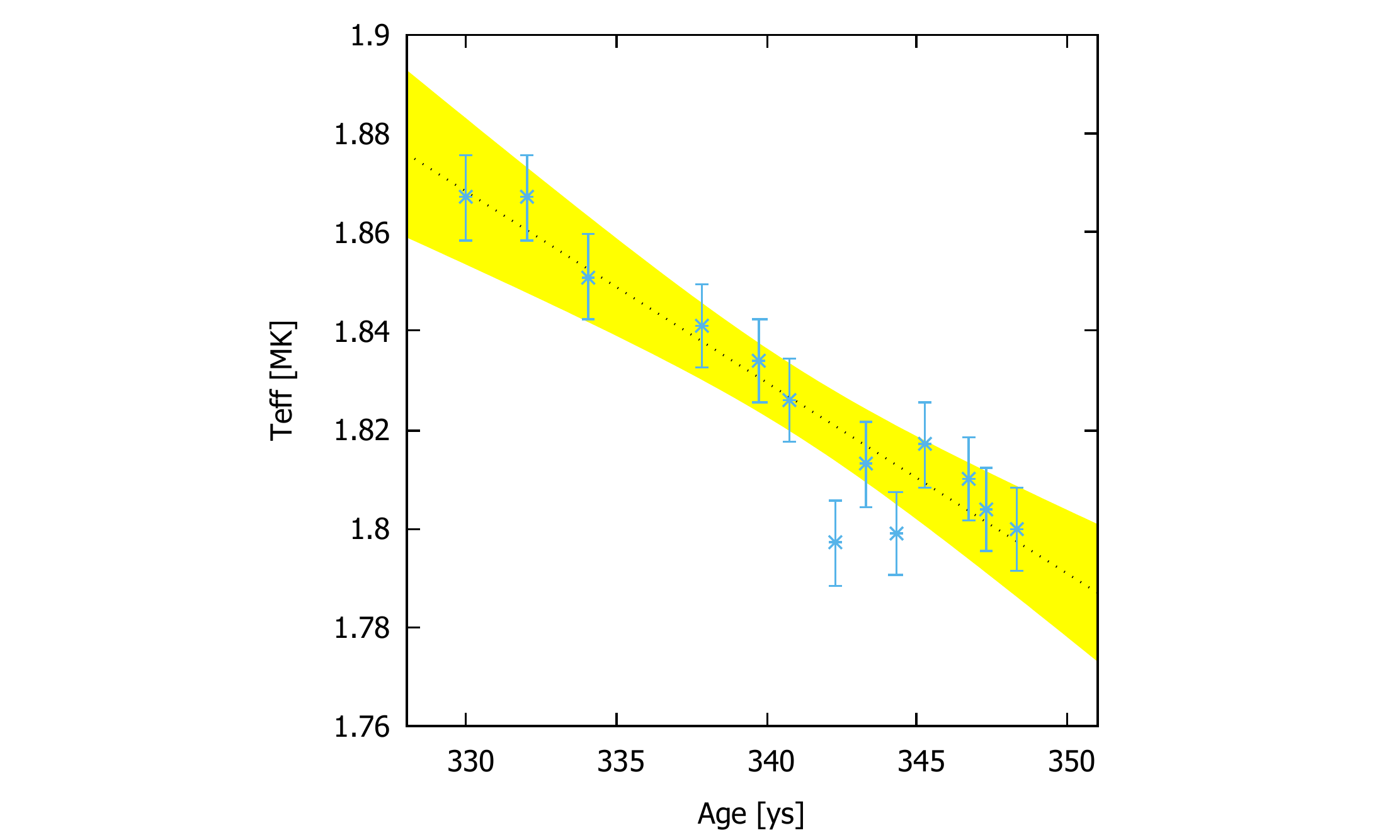}
\end{center}
\caption{(Color online) Surface temperature of the neutron star in the
Cassiopeia A supernova remnant over the past 18 years. Error bars are $1%
\protect\sigma $. Dotted line shows linear fit to the set of $T_{s}$. Yellow
area shows 99\% confidence interval for the surface temperature.}
\label{fig:data}
\end{figure}
%%%%%%%%%%%%%%%%%% 

The next 10 years made significant adjustments to the observational data.
The new work \cite{E13,P13} on the observation of Cas A showed that the
above-mentioned rapid cooling of NS Cas A is not so obvious due to
systematic errors inherent in observations and associated with the problems
of calibrating the detectors of the Chandra satellite telescope. Modern
analysis \cite{PP18,W19} yields upper limits corresponding to 3.3\% or 2.4\%
temperature decreases in 10 years depending on values of the absorbing
hydrogen column density. Although the stellar cooling rate remains high, it
turns out to be significantly less than the previously declared one and fits
well into the scenario described above of neutrino cooling due to PBF
processes. Figure \ref{fig:data} depicts the surface temperature of a
neutron star in a supernova remnant Cassiopeia A over the past 18 years, as
reported in \cite{W19}. The dotted line represents standard least square
linear fit showing an average yearly temperature change rate and yellow area
shows 99\% confidence interval for the Cas A surface temperature\footnote{%
The 99\% confidence interval is constructed under the assumption of a linear
time-dependence of the temperature, which approximately takes place on the
considered part of the NS cooling curve.}.

It can be argued that during the observation period, the effective surface
temperature of CasA should be within the yellow shaded area shown in the
Figure \ref{fig:data} with a 99\% probability. This conclusion can be used
to revise the previous estimate of the strength of the axion-nucleon
interaction, taking into account the new analysis of observational data.
Indeed, the observed rapid cooling of the Cassiopeia A neutron star is
caused by powerful neutrino losses due to PBF processes in the superfluid
neutron core. The more powerfull energy losses from the PBF processes the
more steep decline of the cooling curve. If now, in addition to the neutrino
losses, we take into account the emission of axions in the same PBF
processes, the decline steepness of the cooling curve will depend on the
intensity of the axion emission, that is, on the axion decay constant, which
we are interested in. Since the axion-neutron coupling is inversely
proportional to the axion decay constant $f_{a}$, its lower value should be
limited by the condition that the cooling curve is still in the yellow
region.

In this paper, we study the cooling of NS Cas A due to simultaneous neutrino
and axion energy losses in order to estimate the limit on the axion decay
constant based on the above criterion.

Let us remind that axions are hypothetical Nambu-Goldstone-bosons associated
with the spontaneously broken Peccei-Quinn symmetry that have been suggested
as a solution to the strong-CP violation problem in QCD \cite{P77,W78,Wl78}
but the scale of symmetry-breaking, which is also called the axion decay
constant $f_{a}$, is left undetermined in the theory.

Axions are a plausible candidate for the cold dark matter of the universe,
and a reasonable estimate of the axion decay constant represents much
interest. Though axions arise as Nambu-Goldstone bosons and thus must be
fundamentally massless their interaction with gluons induces their mixing
with neutral pions. Axions thereby acquire a small mass which is inverse
proportional to the scale of symmetry-breaking \cite{BT78,KSS78,PBY87,GKR86}:%
\begin{equation}
m_{a}=0.60\,\text{eV}\,\frac{10^{7}\text{GeV}}{f_{a}}.  \label{ma}
\end{equation}%
We use natural units, $\hbar =c=k_{B}=1$.

Numerous laboratory experiments, as well as astrophysical arguments, were
used to constrain the permissible range of the axion mass $m_{a}$ (see e.g. 
\cite{VZKC,M88,M89,omy96}). Currently \cite{AS83,DF83}, cosmological
arguments give the lower limit $m_{a}>10^{-5}$ eV in order to avoid an
"overclosed universe". The most stringent upper limits on the axion mass
derive from astrophysics. The strength of axion coupling with normal matter
and radiation is limited by the condition that the lifetime of stellar
evolution or the rate of energy loss do not contradict observations. In the
physics of supernova explosions, where the dominant process of energy loss
is the emission of pairs of neutrinos and axions in nucleon bremsstrahlung 
\cite{Br88,Bu88,R93,hpr01}, the requirement that stars do not lose too much
energy due to the emission of axions leads to a lower limit for the
Peccei-Quinn scale $f_{a}$ or, which is the same, up to the upper limit of
the axion mass $m_{a}$. The limit from Supernova 1987A gives $m_{a}<0.01$ eV 
\cite{RS91,J96}. The thermal evolution of cooling neutron stars including
the axion emission in addition to neutrino energy losses was studied in
refs. \cite{i84,U97,s16,s19}. The authors propose upper limits for the axion
mass of the order of $m_{a}<0.06-0.3$ eV, comparing the theoretical curves
with the ROSAT observational data for three pulsars: PSR 1055-52, Geminga,
and PSR 0656 + 14. A similar analysis of the time evolution of the hot young
neutron star in supernova remnant HESS J1731-347 was carried out in works 
\cite{brpr18,L19}\footnote{%
One should remark that the analysis in Refs. \cite{brpr18,L19} was based on
a wrong age estimate for the supernova remnant from \cite{tian08}: the
authors assumed an age a neutron star in SNR HESS J1731-347 of 27 kyr, but
recent studies have revealed that it should be almost an order of magnitude
younger \cite{acero15,cui16,maxted18}.}. Recently, an estimate of the limit
on the decay constant of an axion from the cooling neutron star in
Cassiopeia A was reported in \cite{hnyz18}. We will later compare the
results of this work with our estimate.

Two types of axion models are known: the Kim-Shifman-Weinstein-Zakharov
(hadronic) -- KSVZ model \cite{KSVZ,KSVZ1}, where the axion interacts only
with photons and hadrons, and the Dean-Fischler-Srednitsky-Zhitnitsky --
DFSZ model \cite{DFSZ,DFSZ1} involving the additional axion coupling to the
charged leptons. For a general review on axion physics see, e.g., \cite%
{Kim,Cheng}. The axion phenomenology, in particular in relation with the
astrophysical processes, is largely discussed in \cite{R,R90,T90,R99,R08}.

\section{Energy losses}

\label{sec:losses} Numerical simulations of the Cas A NS cooling are based\
on the public code NSCool \cite{dp89,dp16} which I have modified to include
additional energy losses via the axion emission. I have also introduced
important corrections taking into account the axial anomalous contribution
to the neutrino emissivity caused by the pair breaking and formation (PBF)
processes in the neutron triplet superfluid. Recall that the current version
of the NSCool code incorporates all the corresponding neutrino cooling
reactions: DU, MU, PBF, and bremsstrahlung, but the emissivity of the
neutron $^{3}$P$_{2}$ superfluid requires serious correction. Namely, the
NSCool code includes only complete collective suppression of neutrino
emission in the vector channel\footnote{%
Dipole radiation in the vector channel of weak interactions is absent in a
collision of identical particles.}, as was proved in \cite{LP06,sr9}.
However, the public version of the code does not includes the collective
correction due to anomalous terms in the axial channel\footnote{%
As reported in \cite{bhsp18}, the axial anomalous contribution from ref. 
\cite{L10} is included to the PBF emissivity in the modern (not public)
version of the code.} which additionally significantly reduces the PBF
neutrino emissivity \cite{L10} (for recent review see \cite{L18}). Recall
that PBF processes in a superfluid medium include, in addition to the usual
terms due to the production and absorption of particle-hole pairs, also
anomalous terms describing neutrino emission caused by the production and
absorption of two particles or two holes. This is a very important
correction that, as will be seen later, makes it possible to correctly
describe the observed rate of change in the Cas A NS temperature. \ 

Since the neutrino emission occurs mainly owing to neutron spin
fluctuations, the part of the interaction Hamiltonian relevant for PBF
processes is:%
\begin{equation}
\mathcal{H}_{\nu n}=-\frac{G_{F}C_{A}}{2\sqrt{2}}\delta _{\mu i}\left( \Psi
^{+}\hat{\sigma}_{i}\Psi \right) l^{\mu },  \label{Hnu}
\end{equation}%
Here $l^{\mu }=\bar{\nu}\gamma ^{\mu }\left( 1-\gamma _{5}\right) \nu $ is
the neutrino current, $G_{F}=1.166\times 10^{-5}$ GeV$^{-2}$ is the Fermi
coupling constant, $C_{\mathsf{A}}=1.26$ is the axial-vector coupling
constant of neutrons, and $\hat{\sigma}_{i}$ are the Pauli spin matrices.

The correct form of the PBF neutrino emissivity of the $^{3}$P$_{2}$
superfluid neutrons, as derived in \cite{L10}, reads \ 
\begin{equation}
Q_{\bar{\nu}\nu }^{\mathrm{PBF}}\simeq \frac{2}{15\pi ^{5}}G_{F}^{2}C_{%
\mathsf{A}}^{2}p_{Fn}m_{n}^{\ast }\mathcal{N}_{\nu }T^{7}F_{4}\left(
T/T_{c}\right) ~,  \label{Qnu}
\end{equation}%
where $p_{Fn}$ is the Fermi momentum of neutrons, $m_{n}^{\ast }\equiv
p_{Fn}/V_{Fn}$ is the neutron effective mass, and $\mathcal{N}_{\nu }=3$ is
the number of neutrino flavors; the function $F_{l}$ is given by 
\begin{equation}
F_{l}\left( T/T_{c}\right) =\int \frac{d\mathbf{n}}{4\pi }\frac{\Delta _{%
\mathbf{n}}^{2}}{T^{2}}\int_{0}^{\infty }dx\frac{z^{l}}{\left( \exp
z+1\right) ^{2}},  \label{F0}
\end{equation}%
with $z=\sqrt{x^{2}+\Delta _{\mathbf{n}}^{2}/T^{2}}$. The superfluid energy
gap\footnote{\label{fn4}Note that the definition of the gap amplitude in Eq.
(\ref{Dn}) matches what is implemented in the NSCool code and differs from
the gap definition used in refs. \cite{L10,L18} by $1/\surd 2$ times.}%
\begin{equation}
\Delta _{\mathbf{n}}\left( \theta ,T\right) =\ \Delta \left( T\right) \sqrt{%
1+3\cos ^{2}\theta },  \label{Dn}
\end{equation}%
is anisotropic. It depends on polar angle $\theta $ of the quasiparticle
momentum and temperature $T$.

In standard physical units Eq. (\ref{Qnu}) takes the form 
\begin{eqnarray}
Q_{n\nu }^{\mathrm{PBF}} &=&\frac{4G_{F}^{2}p_{Fn}m_{n}^{\ast }}{15\pi
^{5}\hbar ^{10}c^{6}}\left( k_{B}T\right) ^{7}\mathcal{N}_{\nu }\frac{C_{%
\mathsf{A}}^{2}}{2}F_{4}\left( \frac{T}{T_{c}}\right)   \nonumber \\
&=&1.170\,\times 10^{21}\frac{m_{n}^{\ast }}{m_{n}}\frac{p_{Fn}}{m_{n}c}%
T_{9}^{7}\mathcal{N}_{\nu }\frac{C_{\mathsf{A}}^{2}}{2}F_{4}\left( \frac{T}{%
T_{c}}\right) ~\ \frac{\mathrm{erg}}{\mathrm{cm}^{3}\mathrm{s}}~  \label{QNu}
\end{eqnarray}%
with $T_{9}=T/10^{9}\mathrm{K}$ and $m_{n}$ being the bare neutron mass.
Notice, the neutrino emissivity, as indicated in Eq. (\ref{Qnu}), is 4 times
less than that implemented in the public NSCool code.

The axion interaction with fermions $j$ has a derivative structure. We will
focus in the axion interaction with non-relativistic nucleons. The
corresponding Hamiltonian density can be written in the form: 
\begin{equation}
\mathcal{H}_{an}=\frac{c_{N}}{2f_{a}}\delta _{\mu i}\left( \Psi ^{+}\hat{%
\sigma}_{i}\Psi \right) \partial ^{\mu }a,  \label{Han}
\end{equation}%
where $\Psi $ is a nucleon field, $c_{N}$ is a model dependent numerical
coefficient. For nucleons, the dimensionless couplings $c_{N}$ are related
by generalized Goldberger-Treiman relations to nucleon axial-vector current
matrix elements. A recent determination using lattice QCD finds \cite%
{chvv16,P16}: 
\begin{eqnarray}
c_{n}^{\mathrm{KSVZ}} &=&-0.02(3),~~c_{p}^{\mathrm{KSVZ}}=-0.47\left(
3\right) ,  \nonumber \\
c_{n}^{\mathrm{DFSZ}} &=&0.254-0.414\sin ^{2}{\beta }\pm 0.025,~~  \nonumber
\\
c_{p}^{\mathrm{DFSZ}} &=&-0.617+0.435\sin ^{2}{\beta }\pm 0.025,
\label{cpcn}
\end{eqnarray}%
where $\tan {\beta }$ is the ratio of the vacuum expectation values of the
two Higgs fields in the DFSZ model. Note, a cancellation in the coupling to
neutrons is still possible for special values of $\tan \beta $. In numerical
simulations, the value $\tan \beta =10$~will be used.

A comparison of Eqs. (\ref{Hnu}) and (\ref{Han}) shows that both the
emission of neutrino pairs and axions are caused by fluctuations of the spin
density in the medium. The axion emissivity due to the PBF processes in
neutron spin-triplet superfluid has been derived in ref. \cite{L14} in the
form 
\begin{equation}
Q_{na}^{\mathrm{PBF}}=g_{ann}^{2}\frac{2}{3\pi ^{3}}\frac{p_{Fn}}{m_{n}}%
\frac{m_{n}^{\ast }}{m_{n}}T^{5}F_{2}\left( \frac{T}{T_{c}}\right) ,
\label{Qa}
\end{equation}%
where the function $F_{2}\left( T/T_{c}\right) $ \ is defined in Eq. (\ref%
{F0}). The combination 
\begin{equation}
g_{aNN}=\frac{\mathfrak{c}_{N}m_{N}}{f_{a}}  \label{gann}
\end{equation}%
with $m_{N}$ being the nucleon mass, plays a role of a Yukawa coupling. In
standard physical units, it turns out%
\begin{equation}
Q_{na}^{\mathrm{PBF}}=3.\,\allowbreak 24\times 10^{40}g_{ann}^{2}\frac{p_{Fn}%
}{m_{n}^{\ast }c}\left( \frac{m_{n}^{\ast }}{m_{n}}\right)
^{2}T_{9}^{5}F_{2}\left( \frac{T}{T_{c}}\right) ~\frac{\mathrm{erg}}{\mathrm{%
cm}^{3}\mathrm{s}}  \label{QAn}
\end{equation}

The axion emissivity of the PBF processes in superconducting proton
component is given by%
\begin{equation}
Q_{pa}^{\mathrm{PBF}}=1.\,\allowbreak 55\times 10^{40}g_{app}^{2}
\left( \frac{m_{p}^{\ast }}{m_{p}}\right)
^{2}T_{9}^{5}\left( \frac{p_{Fp}}{m_{p}c}\right) ^{3}\frac{6}{7}F_{2}\left( 
\frac{T}{T_{cp}}\right) ~\frac{\mathrm{erg}}{\mathrm{cm}^{3}\mathrm{s}}
\label{QAp}
\end{equation}%
It differs from the proton PBF emission of neutrino pairs in the axial
channel (see e.g. \cite{kv08}) only in the coupling constant and phase volum
of freely escaping particles resulting in the weaker temperature dependence.

\begin{figure}[tbp]
\begin{center}
\includegraphics[width =1\textwidth]{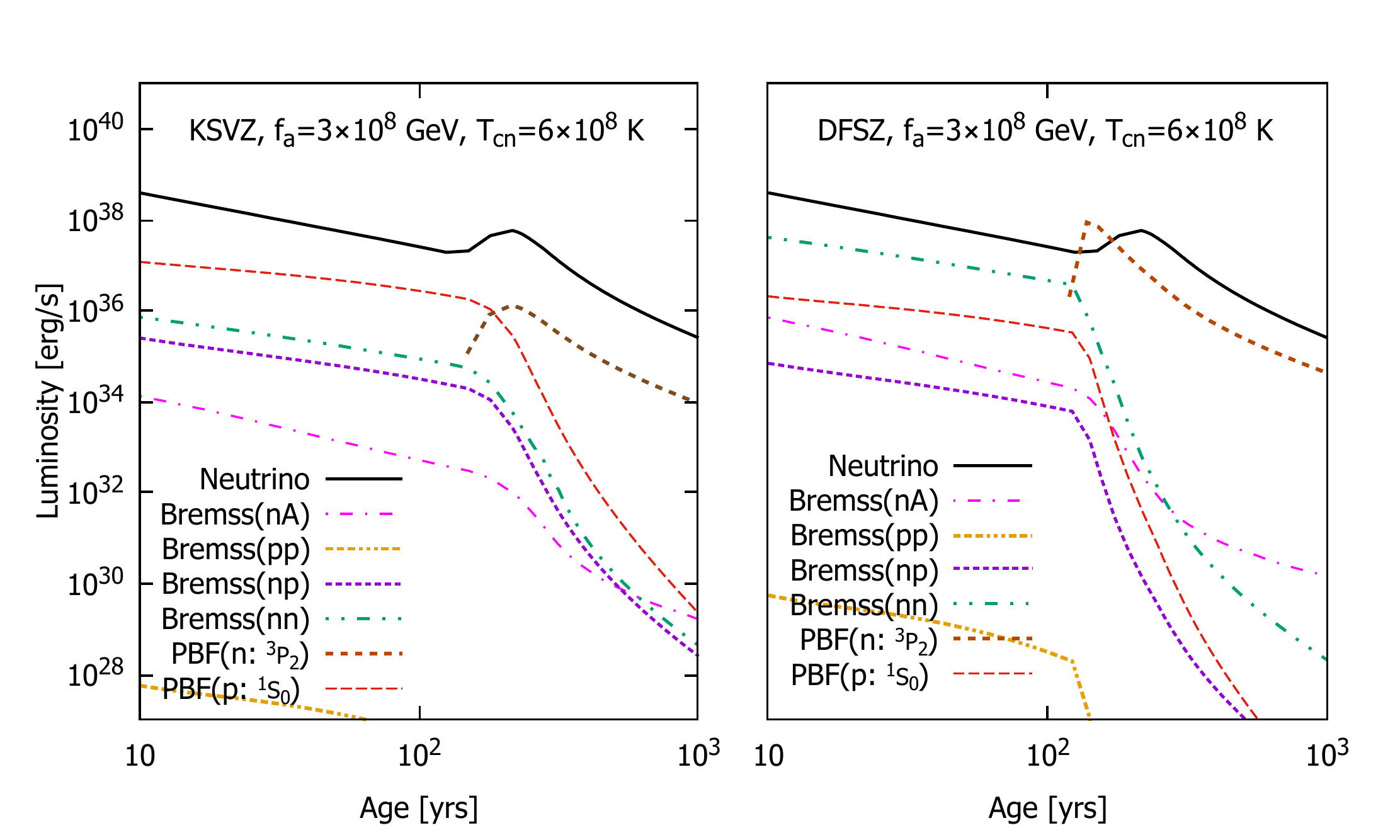}
\end{center}
\caption{(Color online) Luminosity of each axion emission and the total
neutrino emission processes as a function of time. Dimensionless coupling
constants $c_{N}$ are as indicated in Eqs. (\protect\ref{cpcn}) and $\tan 
\protect\beta =10$.}
\label{fig:lum}
\end{figure}

Axion-nucleon couplings also cause the emission of axions through NN
bremsstrahlung processes and modified urca processes, which have so far been
studied in the literature (see, for example, \cite%
{fm79,i84,hpr01,sjg04,Y01,brpr18} and references therein ). We omit explicit
expressions for these processes for brevity. In Fig. \ref{fig:lum}, we show
luminosities of various axion emission processes in the KSVZ and DFSZ models
with $f_{a}=3\times 10^{8}$ GeV as functions of time. For comparison, we
also show the total luminosity of neutrino emission.%
%%%%%%%%%%%%%%%%%%
%%%%%%%%%%%%%%%%%% 

It is instructive to compare this figure with Fig. 1 of ref. \cite{hnyz18}
where the same calculation was carried out. In the graphs presented in this
work, the surprisingly small contribution of the PBF processes in the
neutron spin-triplet superfluid to axion losses immediately catches the eye,
while these processes dominate in neutrino losses after the neutron
superfluidity onset in the NS core. The authors do not provide an explicit
analytical expression for the corresponding axion emissivity, only referring
to the work \cite{L14}. Let me remind you that in this work the same
equation (\ref{Qa}) is derived which is used in the present work (but see
footnote \ref{fn4}).

\section{Cooling simulation}

\label{sec:sim}

Following \cite{w15} I consider a non-rotating neutron star of a mass $%
M_{NS}=1.441M_{\odot }$ with Fe envelope. The equilibrium structure of the
star was obtained as a solution to the relativistic
Tolman-Oppenheimer-Volkov equations supplemented by the BSk21 equation of
state \cite{f13,p13}.%%%%%%%%%%%%%%%%%%

\begin{figure}[tbp]
\includegraphics[width =1\textwidth]{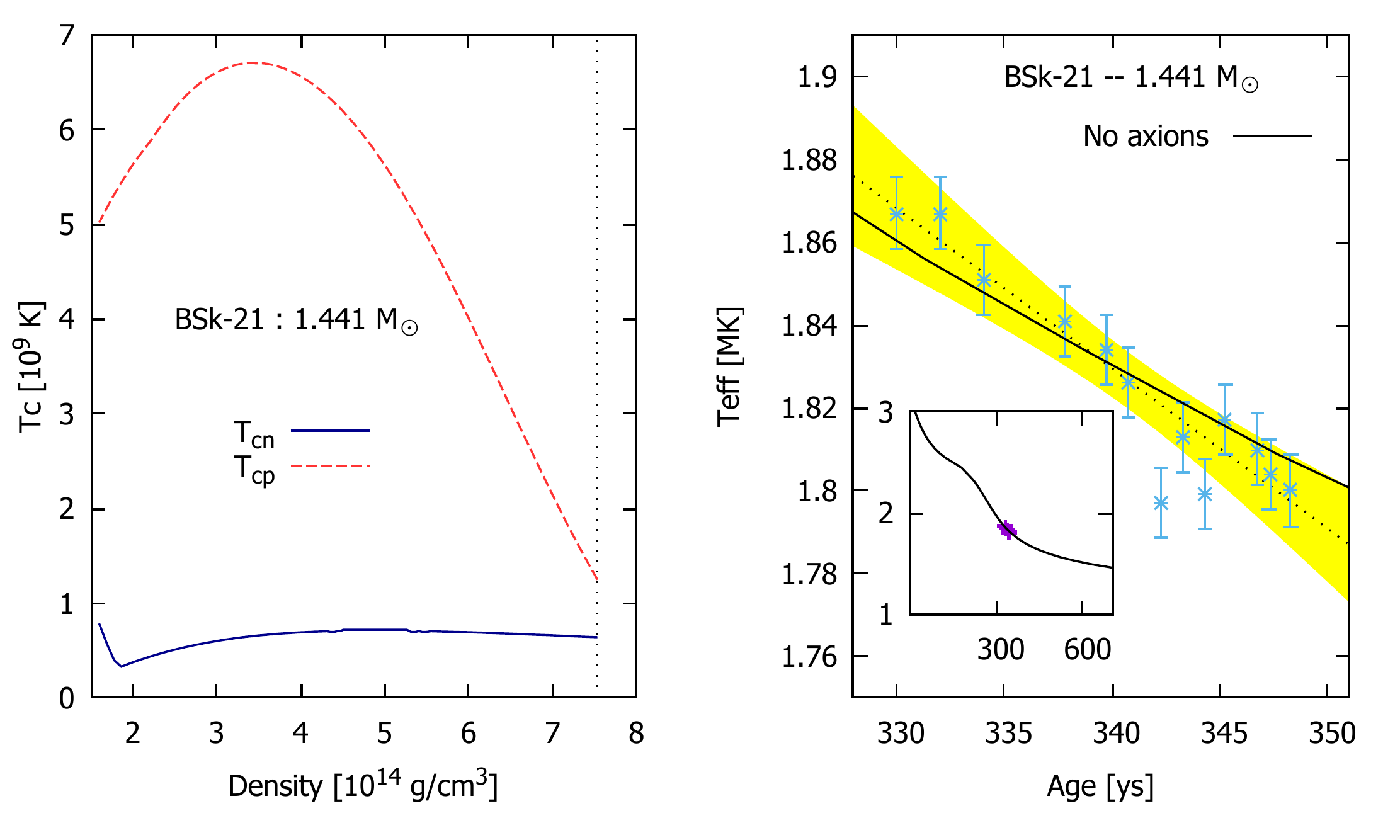}
\caption{(Color online) Left panel: Critical temperatures $T_{c}$ for proton
singlet superfluidity and for triplet superfluidity of neutrons in the NS
core as a function of a mass density constructed using the BSk21 ($%
M_{NS}=1.441M_{\odot },\ R=12.6\ \mathsf{km}$). Right panel: Surface
temperature $T_{s}$ without redshift as a function of age for $1.441M_{\odot
}$ BSk21 NS with an iron envelope. The black line shows the temperature
change obtained in the minimum cooling scenario. The inset shows the same NS
cooling trajectory for a longer time.}
\label{fig:tcts}
\end{figure}
%%%%%%%%%%%%%%%%%% 

As was found in ref. \cite{St,P} the minimal cooling scenario puts stringent
constraints on the temperature $T_{cn}$ for the onset of neutron
superfluidity in the Cas A NS. Namely, the transition temperature dependence
on the density should have a wide peak with maximum $T_{cn}(\rho )\approx
(5-8)\times 10^{8}$~K. For the NS cooling simulations, I use the widely
adopted CCDK model \cite{CCDK,e96} for the \ proton gap and the TToa model 
\cite{tt4} for the neutron gap in the NS core. For singlet pairing of
neutrons I have chosen the "SFB" model \cite{sfb03}. The choice of other
models for the singlet pairing of neutrons slightly affects the result,
since the $^{1}$S$_{0}$ pairing of neutrons occurs only in the inner NS
crust. Critical temperatures for proton singlet superfluidity and for
triplet superfluidity of neutrons in the NS core vs a mass density are shown
in the left panel of Fig. \ref{fig:tcts}. The accepted model of proton
superfluidity assumes a sufficiently high critical temperature, which is
necessary for sufficient suppression of modified Urca processes prior to the
current era of rapid cooling.

Let us first consider NS cooling without axion emission. The corresponding
cooling trajectory obtained as a result of numerical simulation is shown in
right panel of Fig. \ref{fig:tcts} with a black line. The line demonstrates
temporal behavior the NS surface temperature. The cooling curve describes
the observed decrease in temperature and is within the confidence interval.

\begin{figure}[tbp]
\includegraphics[width =1\textwidth]{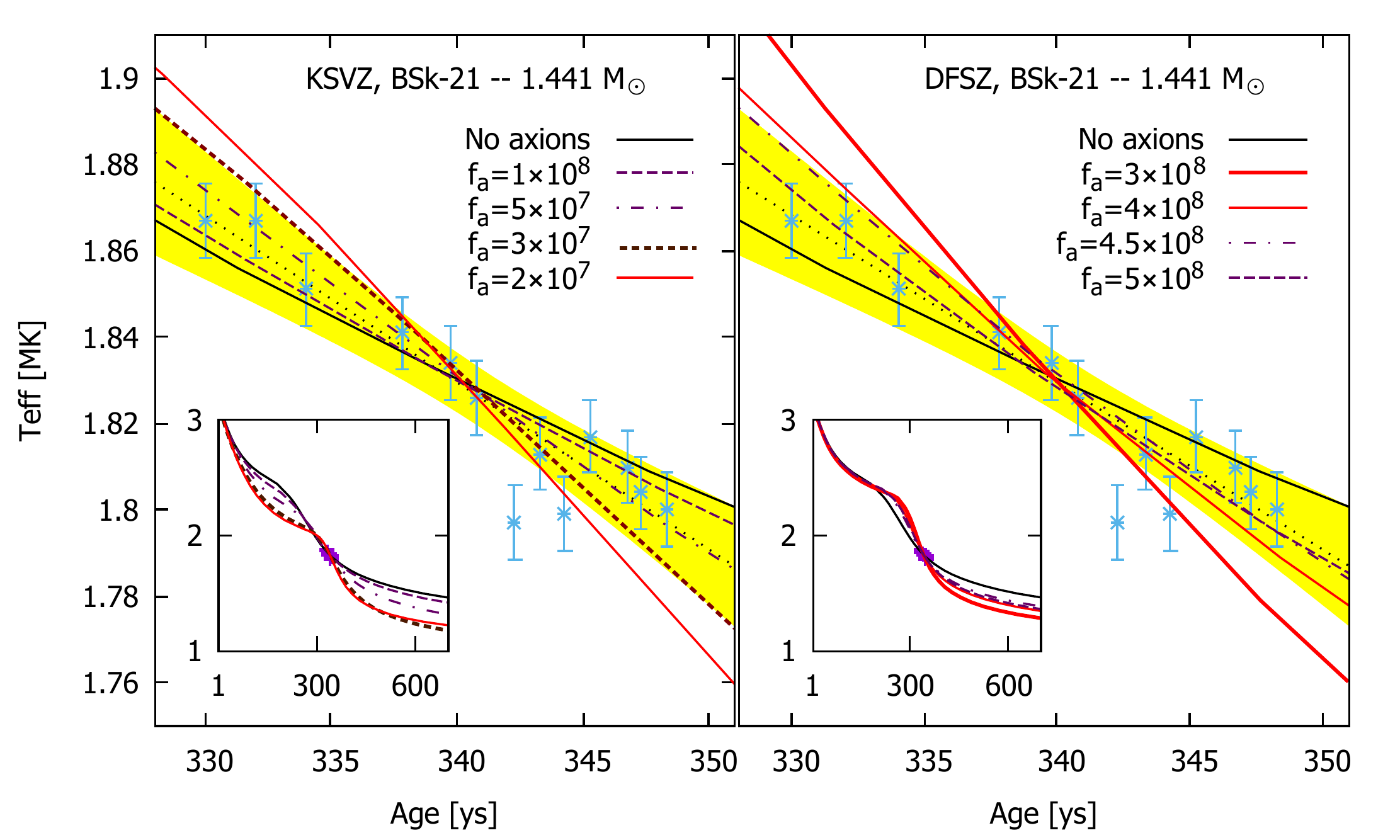}
\caption{(Color online) Cooling curves compared to observed data. The inset
shows the same NS cooling trajectories for a longer time.}
\label{fig:axion}
\end{figure}

Let us now investigate the influence of axion emission processes on the
evolution of the NS temperature by adding axion energy losses. Taking the
dimensionless coupling constants $c_{N}$ given in Eq. (\ref{cpcn}) with $%
\tan \beta =10$, I will use the axion decay constant $f_{a}$ as a parameter.
The best-fit curves of the surface temperature $T_{s}$ for several values of 
$f_{a}$ are shown in Fig. \ref{fig:axion} for the KSVZ and DFSZ models
together with the trajectory obtained in of the minimal neutrino cooling
scenario. %%%%%%%%%%%%%%%%%%

%%%%%%%%%%%%%%%%%% 
For each curve, I vary the neutron triplet gap parameters and the amount of
light elements in the NS envelope (the envelope parameter $\eta $) to fit
the observed data. As one can see, as $f_{a}$ decreases, the NS surface
temperature falls steeper during the observation period, and, finally, below
a certain critical value $f_{a}^{\mathrm{cr}}$, it goes beyond the 99\%
confidence interval. With the baseline scenario of Cas A NS cooling, we can
now definitely constrain the emission of axions by the condition that the
cooling curve should be localized in the yellow region. The KSVZ axion model
yields $f_{aKSVZ}^{\mathrm{cr}}=3\times 10^{7}$ GeV, while in the DFSZ model
one gets $f_{aDFSZ}^{\mathrm{cr}}=4.5\times 10^{8}$ GeV.

\section{Discussion and conclusion}

\label{sec:disc}

The influence of emission of axions by nucleons on the steepness of the
temperature decline of the NS surface is investigated. Additional axion
radiation accompanying the PBF neutrino emission of nucleons in the NS core
significantly increases the NS cooling rate, making the cooling trajectory
steeper. At some critical value of the axion decay constant, the best fit
cooling trajectory goes beyond the 99\% confidence interval obtained from
the observational data. This condition is used to restrict the minimum value
of the axion decay constant to obtain $f_{aKSVZ}>3\times 10^{7}$ GeV and $%
f_{aDFSZ}>4.5\times 10^{8}$ GeV.

Note that in the case of the KSVZ model, the above estimate contradicts the
result obtained in \cite{hnyz18}, and agrees well with the estimates
obtained in \cite{L14,s16,brpr18}. Remind that ref. \cite{L14} sets a
restriction $f_{aKSVZ}\gtrsim 5\times 10^{7}$ GeV for $c_{n}=-0.02$, while
ref. \cite{s16} sets $f_{aKSVZ}\gtrsim \left( 5\div 10\right) \times 10^{7}$
GeV for the KSVZ model. A similar estimate was obtained in \cite{brpr18} : $%
f_{aKSVZ}>6.7\times 10^{7}$ GeV. For the DFSZ model, the authors of \cite%
{brpr18} state: $g_{ann}^{2}=7.7\times 10^{-20}$. If one replaces $c_{n}$ in
this ratio with the value specified in the formula (\ref{cpcn}) with $\tan
\beta =10$, as I use in my calculations, it is easy to find that $%
f_{aDFSZ}>5.\,\allowbreak 2\times 10^{8}$ GeV, in good agreement with the
result obtained in the present work.

Unfortunately, the coupling constants $c_{N}$ depend on the axion model.
Given the QCD uncertainties of the hadronic axion models \cite{k85,s85,g82},
the dimensionless constant $c_{n}$ could range from $-0.05$ to $0.14$. While
the canonical value $c_{n}^{KSVZ}=-0.02$ is often used as generic examples,
in general a strong cancelation of $c_{n}^{KSVZ}$ below $c_{n}^{KSVZ}=-0.02$
is also allowed. In case of $c_{n}^{KSVZ}\rightarrow 0$ a powerfull PBF
emission of KSVZ axions from $^{3}$P$_{2}$ neutron pairing is impossible.

A few comments should also be made regarding the observed cooling rate of
Cassiopea A remnant, which is still controversial. In works\ \cite{P13, PP18}
the authors state that the previously described rapid cooling of NS Cas A is
probably a systematic artifact, and they cannot rule out the standard slow
cooling for this NS. Their results (2006-2012) are consistent with no
temperature drop at all, or less temperature drop than previously reported,
although the associated uncertainties are too large to firmly rule out the
previously reported rapid cooling. Further observations are needed to more
accurately estimate the rate of temperature drop. Note, however, that the
theoretical cooling trajectory shown in Fig. \ref{fig:tcts} is in even
better agreement with slower cooling. In this case the linear regression fit
(dotted line) would be closer to the theoretical neutrino cooling trajectory
which should result to a more strong restrictions to the axion decay
constant.

\end{document}